\documentclass[journal,twoside,web]{ieeecolor}
\usepackage{lcsys}
\usepackage{cite}
\usepackage{amsmath,amssymb,amsfonts}
\usepackage{graphicx}
\usepackage{textcomp}
\usepackage{threeparttable}
\usepackage{hyperref}
\usepackage{amssymb}
\usepackage{tabu}
\usepackage[utf8]{inputenc}
\usepackage{setspace}
\usepackage{amsmath}
\usepackage{amsfonts}
\usepackage{subcaption}
\usepackage{multirow}
\usepackage{tabularx}
\usepackage{array}
\usepackage{makecell}
\usepackage{algpseudocode}
\usepackage{dblfloatfix}
\newtheorem{theorem}{\textbf{Theorem}}
\newtheorem{lemma}{\textbf{Lemma}}
\newtheorem{fact}{\textbf{Fact}}
\usepackage{authblk}
\usepackage{float}
\usepackage[table]{xcolor}
\newtheorem{definition}{\textbf{Definition}}
\newtheorem{proposition}{\textbf{Proposition}}
\usepackage{color}
\usepackage{tikz}  
\newcommand{\absElem}[1]{|#1 |_{\text{abs}}}

\DeclareMathOperator{\relu}{ReLU}

\def\BibTeX{{\rm B\kern-.05em{\sc i\kern-.025em b}\kern-.08em
    T\kern-.1667em\lower.7ex\hbox{E}\kern-.125emX}}
\begin{document}
\title{Ensuring Both Positivity and Stability Using Sector-Bounded Nonlinearity for Systems with Neural Network Controllers}
\author{Hamidreza Montazeri Hedesh$^1$ and  Milad Siami$^1$ \IEEEmembership{Senior Member, IEEE}
\thanks{This material is based upon work supported in by grants ONR N00014-21-1-2431, NSF 2121121, NSF 2208182, the U.S. Department of Homeland Security under Grant Award Number 22STESE00001-03-02, and by the Army Research Laboratory under Cooperative Agreement Number W911NF-22-2-0001. The views and conclusions contained in this document are solely those of the authors and should not be interpreted as representing the official policies, either expressed or implied, of the U.S. Department of Homeland Security, the Army Research Office, or the U.S. Government.}
\thanks{$^1$H. Montazeri Hedesh and M. Siami are with the Department of Electrical \& Computer Engineering, Northeastern University, Boston, MA 02115, USA.
(e-mails: {\tt\footnotesize \{montazerihedesh.h, m.siami\}@northeastern.edu}).}}

\maketitle
\thispagestyle{empty}
\begin{abstract}
This paper introduces a novel method for the stability analysis of positive feedback systems with a class of fully connected feedforward neural networks (FFNN) controllers. By establishing sector bounds for fully connected FFNNs without biases, we present a stability theorem that demonstrates the global exponential stability of linear systems under fully connected FFNN control. Utilizing principles from positive Lur'e systems and the positive Aizerman conjecture, our approach effectively addresses the challenge of ensuring stability in highly nonlinear systems. The crux of our method lies in maintaining sector bounds that preserve the positivity and Hurwitz property of the overall Lur'e system. We showcase the practical applicability of our methodology through its implementation in a linear system managed by a FFNN trained on output feedback controller data, highlighting its potential for enhancing stability in dynamic systems.
\end{abstract}

\begin{IEEEkeywords}
Neural Networks, Neural Network Verification, Positive Systems, Neural Network Bound, Stability Analysis
\end{IEEEkeywords}

\allowdisplaybreaks
\section{Introduction}
\IEEEPARstart{M}{ultilayer} feedforward neural networks are universal approximators \cite{HORNIK1989359}. This fact has led to extensive applications of neural networks (NN) across various fields. In the field of control systems, there has been  a longstanding interest in using NNs as controllers in the feedback loop of the systems. However, many challenges have emerged along the way \cite{sznaier2022role}. NN-controllers inherit many challenges due to their complex and highly nonlinear structure. Input sensitivity, lack of robustness, and lack of stability certificates are some of the issues that pose severe risks in safety-critical systems. Consequently, many recent studies have tried to address these shortcomings and propose methods for verification of NNs in closed loop.

Control theory provides an arsenal of valuable tools conducive to the verification of NN-controllers, such as Lyapunov functions, robust control techniques, passivity analysis, and Control Barrier Functions (CBFs), among others. Researchers are actively exploring the integration of these tools to develop robust verification methods. In a notable instance, the studies \cite{yin2021stability, yin2021imitation} leverage Quadratic Constraints (QCs), combining them with Lyapunov functions for a comprehensive verification approach. The authors continued their work on Integral Quadratic Constraint (IQC) for verification of NNs in \cite{Gu_Yin_Ghaoui_Arcak_Seiler_Jin_2022}. The authors of \cite{fazlyab2021introduction,hashemi2021certifying} employ QCs along with the S-procedure to establish stability conditions and address reachability concerns. Some studies like \cite{jin2020stability} used passivity theorem. The authors in \cite{qin2021learning} propose the use of CBF to address these challenges, presenting a distinctive perspective. The article \cite{richardson2023strengthened} used Lur'e systems, Circle, and Popov criteria to analyze the stability of such systems. The study \cite{newton2022stability} used the Sum of Squares (SOS) method to device Lyapunov functions for stability assurance of NN-controllers in the feedback loop of nonlinear systems. Many other creative methods have been used to verify NN-controlled systems. \cite{de2023iss,commuri1997cmac,NEURIPS2023_d79c1390}. A comprehensive review of the methodologies used in this domain is presented in \cite{dawson2023safe}.

These mathematical tools are very useful in the analysis of the highly nonlinear structures inherent in NN-controlled systems. However, they carry their limitations such as complexity and lack of scalability to large systems. Some studies have tried to tackle this drawback\cite{gates2023scalable}. For example, \cite{zhang2018efficient,chen2022deepsplit} came up with algorithms to split the verification problem into sub-problems and solve them more efficiently. Meanwhile, the search for new mathematical tools is still ongoing, and there exist many unexplored useful tools \cite{khalil2002grizzle}. Recently, one of the fundamental and simple methods for stability verification of nonlinear systems came into light. Aizerman's conjecture for absolute stability was proven to be wrong universally \cite{bragin2011algorithms}. However, very recently, the conjecture was proved to hold for the class of positive systems \cite{drummond2022aizerman}. These new findings opened the way to developing a simple and scalable method for the verification of NN-controllers. Our findings presented in this paper are based on this method. The use of Lur'e systems in the analysis of NNs is not a new concept, and there is literature covering this method such as \cite{soykens1999lur}. However, the use of the Aizerman conjecture was absent due to the general disprove of it. In this article, we use this simple yet resourceful conjecture for positive NN-controlled systems. 

In this paper, we introduce a sector bound for a fully connected FFNN without biases. The sector bound is based on the properties of the activation function of the NN and its weights. The establishment of sector bounds was necessitated by their indispensability in our verification methodology. Existing bounds such as IBP \cite{gowal2018effectiveness}, Lipschitz constant \cite{szegedy2013intriguing}, and Quadratic Constraints as described in \cite{yin2021stability,fazlyab2021introduction} lacked the requisite structure for verification via the Aizerman conjecture. Consequently, we devised our own bounds. Our proposed bound diverges significantly from the IBP method and the Lipschitz constant of the NN. The IBP method involves propagating the input through layers, identifying bounds for the output polytopes of each layer. However, the resulting output bound is a set that lacks explicit connection with the input to the NN. Additionally, the Lipschitz constant, being a scalar, may even exceed the sector bound of a given nonlinear function. In contrast, sector bounds establish a direct relationship between the output and input, potentially spanning higher dimensions than a scalar. These characteristics make sector bounds particularly valuable for nonlinear analysis of NNs, notably in the context of the Aizerman and Kalman conjecture. Previous works such as \cite{yin2021stability} and \cite{fazlyab2021introduction} also utilize sector bounds, albeit restricting the nonlinearity to the NN's activation functions. In this paper, we introduce a sector bound for the entire NN, motivated by its necessity in addressing the Aizerman conjecture.
Using the proposed sector bound, we present our stability theorem.

The contribution of our work is twofold. First, an introduction of a sector bound for fully connected without biases which can be used further in many applications, like forward reachable sets of NNs, or nonlinear control analysis of the closed loop systems. Moreover, we present a simple verification test for the stability of positive closed-loop systems consisting of NN-controllers and linear time-invariant (LTI) systems. The stability condition is simply based on the calculated sector bounds. It only checks that the upper and lower bounds of the system are Metzler and Hurwitz, respectively. The calculation does not require a huge amount of memory and time, making it suitable for extending to more complex systems and system of systems. Another standout point of this research compared to the literature is that it can handle continuous-time systems. Moreover, this paper addresses global asymptototic stability (within the context of positive systems) of NN-controlled systems, while most of the available verification methods study local stability.

\subsection{Notation}
The orders $<,\geq$, when applied to matrices and vectors, act elementwise. The set of real numbers is denoted by $\mathbb R$, and the set of nonnegative real numbers by $\mathbb R_+$. The set of real n-dimensional vectors and the set of real $m \times n$-dimensional matrices are denoted by $\mathbb R^n$ and $\mathbb R^{m\times n}$, respectively. For a real matrix (vector) $Q$, the notation $Q > 0$ indicates that all elements of $Q$ are positive, $Q\geq 0$ indicates that all elements of $Q$ are nonnegative. $<$ and $\leq$ are defined similarly. We define $\mathbb R_+^n := \{\nu \in \mathbb R^n : \nu \geq 0\}$. In addition, we define $\mathbb R^{n\times m}_+$ with obvious modifications. $\mathbf{0}_n$ and $\mathbf{I}_n$ denote a vector of zeros and the Identity matrix of size $n$, respectively. Furthermore, given a matrix $A$, the elementwise absolute value is denoted as $\absElem{A}$, where
\begin{equation*}
\small
\absElem{A} = \begin{pmatrix}
\left|a_{11}\right| & \left|a_{12}\right| & \cdots & \left|a_{1n}\right| \\
\left|a_{21}\right| & \left|a_{22}\right| & \cdots & \left|a_{2n}\right| \\
\vdots  & \vdots  & \ddots & \vdots  \\
\left|a_{m1}\right| & \left|a_{m2}\right| & \cdots & \left|a_{mn}\right|
\end{pmatrix}.
\end{equation*}
With this definition, it is evident that for two real matrices $Q$ and $T$ of compatible dimensions,
$\absElem{QT} \leq \absElem{Q}\absElem{T}$.
\section{Preliminaries}

\subsection{Positive Systems}
 In the world of control theory, the concept of positive systems plays an important role due to their widespread applicability in various domains such as economics, biology, and engineering. Positive systems are distinguished by their intrinsic property that, given nonnegative initial conditions, their state variables remain non-negative for all future times. This section aims to succinctly define positive systems, outline the essential conditions for their positivity, and explain technical terms. Consider the generic linear control system:
\begin{equation}\label{eq:generallti}
    \dot x(t) = A x(t) + B u(t), \quad y(t) = C x(t),
\end{equation}
where $A \in \mathbb R^{n\times n}$, $B \in \mathbb R^{n\times m}$, and $C \in \mathbb R^{p\times n}$ are constant matrices. In addition, $u(t) \in \mathbb R^m$, $x(t) \in \mathbb R^n$, and $y(t) \in \mathbb R^p$ denote input, state, and output variables.

\begin{definition}[\textbf{Positive Linear System}]\label{def:positive system}
A linear system, as characterized by \eqref{eq:generallti}, is defined to be
\emph{Positive} if, for all $t>0$, given initial conditions $x(0) \geq 0$ and input $u(t) \geq 0$, the state $x(t)$ remains non-negative.
\end{definition}

\begin{definition}[\textbf{Metzler Matrix}]
A matrix $M = [m_{ij}] \in \mathbb{R}^{n \times n}$ is defined as a \emph{Metzler matrix} if all its off-diagonal elements are non-negative; that is, $m_{ij} \geq 0$ for all $i \neq j$.
\end{definition}

\begin{fact}\label{fact1}
Consider the linear differential equation $\dot x = Mx.$ The system described above is positive if and only if the matrix $M$ is Metzler. Moreover, if $M$ is also Hurwitz, there must exist a vector $v \in \mathbb{R}_+^n$, $v > 0$ with all positive entries such that $v^T M < 0$ has all negative entries. This fact is adopted from \cite{drummond2022aizerman} and detailed proofs of the claims can be found in the references therein \cite{berman1989nonnegative,haddad2010nonnegative}.
\end{fact}

\begin{proposition}
Given a system described by \eqref{eq:generallti} and considering the definition of a positive system as per Definition \ref{def:positive system}, the system is positive if and only if matrix $A$ is Metzler, and matrices $B \in \mathbb R^{n\times m}_+$ and $C\in \mathbb R^{p\times n}_+$ \cite{ebihara2016analysis}.
\end{proposition}

\subsection{Lur'e Systems and Aizerman's Conjecture}
\begin{figure}[t]
    \centering
    \includegraphics[width=.35\linewidth]{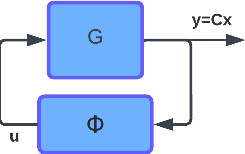}
    \caption{\small Lur'e system with plant $G$ and nonlinear controller $\Phi$.}
    \vspace{-.4cm}
    \label{fig:syslook}
\end{figure}
\begin{figure}[t]
    \centering
            \begin{subfigure}{.20\textwidth}
                \centering
                \includegraphics[width=\linewidth]{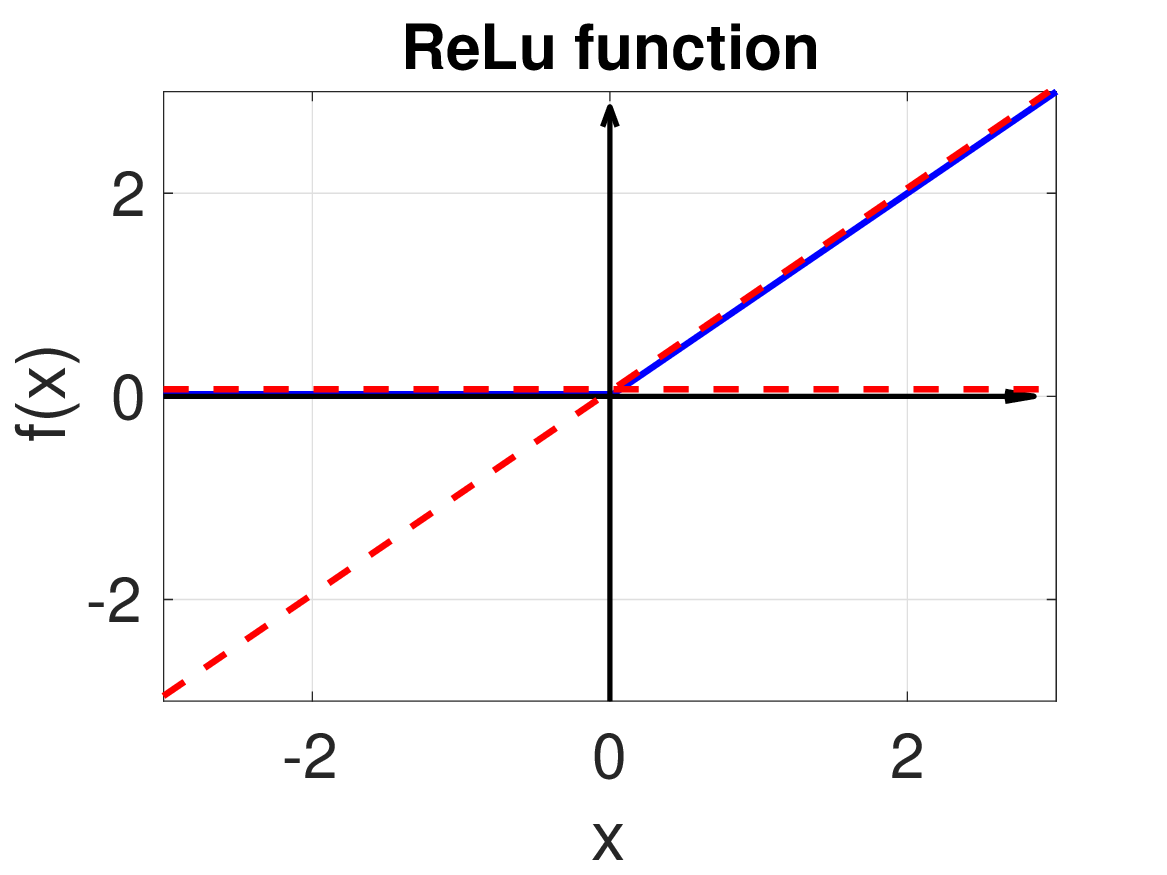} 
                \label{fig:Relusect}
            \end{subfigure}%
            \begin{subfigure}{.20\textwidth}
                \centering
                \includegraphics[width=\linewidth]{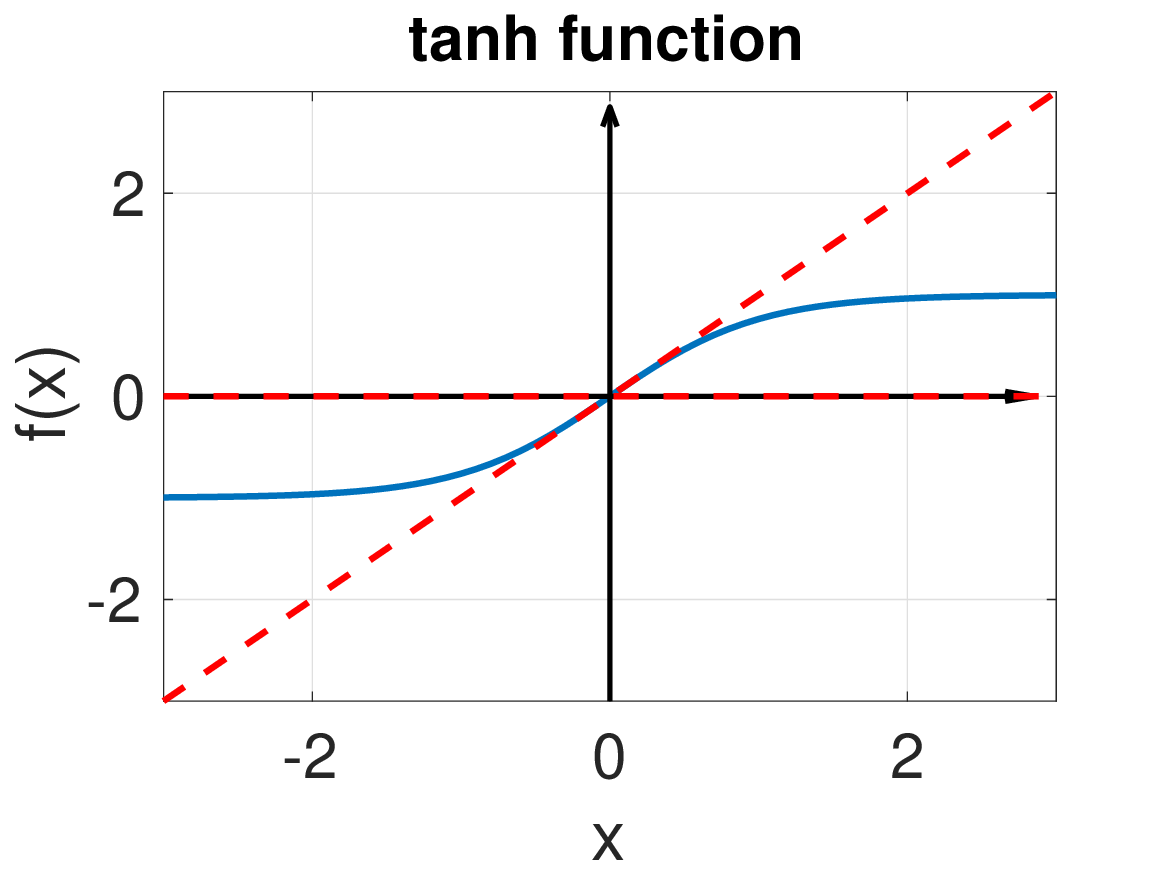}
                \label{fig:tanhsect}
            \end{subfigure}
            \vspace{-.5cm}
    \caption{\small $\relu$ and $\tanh$ sector-bounded in $\left[0,1\right]$.}
        \vspace{-.65cm}
    \label{fig:sectorfigures}
\end{figure}

The classical framework of Lur'e systems and Aizerman's conjecture applies to Single Input Single Output (SISO) systems. These involve a SISO linear system interconnected with a nonlinearity in the feedback loop, as shown in Fig. \ref{fig:syslook}. Consider a closed-loop nonlinear control system described by
$\dot{x}(t) = Ax(t) + b\Phi(\sigma), \quad \sigma = c^Tx(t),$
where $x \in \mathbb{R}^n$ is the state vector, $A \in \mathbb{R}^{n \times n}$ is the system matrix, $b, c \in \mathbb{R}^n$ are constant vectors. The nonlinearity $\Phi: \mathbb{R} \mapsto \mathbb{R}$ satisfies $\Phi(0) = 0$. 
Aizerman's conjecture states that the system is globally asymptotically stable if, for every linearization of \( \Phi \) within a sector \([k_1, k_2]\), where \( k_1 \) and \( k_2 \) are real constants, the linearized system
$\dot{x}(t) = Ax(t) + bk\sigma, \quad \sigma = c^Tx(t), \quad k \in [k_1, k_2],$
is asymptotically stable. The sector condition is
\begin{equation}\label{eq:sisosector}
k_1 \leq \frac{\Phi(\sigma)}{\sigma} \leq k_2, \quad \forall \sigma \neq 0.
\end{equation}
Note that common NN activation functions, such as $\tanh$ and $\relu$, satisfy sector bounds within $[0,1]$. Fig. \ref{fig:sectorfigures} illustrates these functions with the global sector $[0,1]$.
However, exceptions to Aizerman's conjecture exist, with counterexamples showing systems that meet the conjecture's criteria but have both a stable equilibrium and a stable periodic solution.
Nonetheless, under more stringent conditions, such as system positivity, the conjecture's validity is reinforced \cite{drummond2022aizerman}.

\subsection{Positive Lur’e Systems}
While the Aizerman conjecture was traditionally associated with SISO systems, this discussion extends to encompass Multi-Input Multi-Output (MIMO) systems. Consider the class of Lur'e systems shown in Fig. \ref{fig:syslook}.
Assume the interconnection of the linear system \eqref{eq:generallti} and the static nonlinear feedback $u = \Phi(y,.)$. The dynamics of such an interconnection can be succinctly represented by:
\begin{equation}\label{eq:positiveluresystem}
    \dot x = Ax(t) + B \Phi(Cx,t),
\end{equation}
where  $A \in \mathbb R^{n\times n}$, $B \in \mathbb R^{n\times m}$, and $C \in \mathbb R^{p\times n}$ are constant matrices and the multivariate function $\Phi: \mathbb{R}^p \times \mathbb{R} \mapsto \mathbb{R}^m$ is assumed to satisfy $\Phi(0, t) = 0, \forall t \geq 0$, ensuring that $x_* = 0$ serves as an equilibrium point for \eqref{eq:positiveluresystem}. A fundamental assumption in this study is the well-posedness of the feedback interconnection depicted in Fig. \ref{fig:syslook}.
We assume the existence of a unique and locally absolutely continuous function $\chi:\mathbb{R}_+ \mapsto \mathbb{R}^n$ that satisfies equation \eqref{eq:positiveluresystem} almost everywhere, for any initial condition $x(0) \in \mathbb{R}^n$. The well-posedness of the system is predicated on standard conditions applied to the nonlinearity $ \Phi(z,t) $: it must be locally Lipschitz in $ z $ and measurable in $ t $, along with certain mild boundedness conditions. These prerequisites ensure the existence and uniqueness of the solution, as delineated in foundational control theory literature such as 
~\cite[Th. 54, Prop. C.3.8]{sontag2013mathematical}.

Given the MIMO system context, the multivariable function $\Phi$ transcends the traditional sector bound definition as in \eqref{eq:sisosector}.
Within the context of positive systems, defining a sector bound for the multivariable function $\Phi$ is most coherently articulated in terms of componentwise inequalities. For the case of positive systems, we define a sector bound for $\Phi$ as following. Given two matrices $\Sigma_1, \Sigma_2 \in \mathbb{R}^{m \times p}$, with $\Sigma_1 \leq \Sigma_2$\footnote{Sign conventions: This article makes no assumptions about the signs of the sector bounds $\Sigma_1$ and $\Sigma_2$.}, the function $\Phi$ is considered to be within sector $[\Sigma_1,\Sigma_2]$ if:
\begin{equation}\label{eq:mimo sector bound}
\Sigma_1z \leq \Phi(z, t) \leq \Sigma_2z, \quad \forall z \in \mathbb{R}^p_+, \forall t \geq 0.
\end{equation}
By establishing the sector bound within a MIMO system framework, we introduce a lemma from \cite{drummond2022aizerman} that presents a necessary and sufficient condition for the positivity of a Lur'e system. The lemma, provided below, is included here for reference as it becomes essential later in the paper.
\begin{lemma}\label{lem:lemma1}
Consider the Lur'e system as described in \eqref{eq:positiveluresystem}, assuming that both $B, C \geq 0$. The system \eqref{eq:positiveluresystem} is a positive system for any $\Phi \in \text{Sector}[\Sigma_1, \Sigma_2]$ if and only if the matrix $A + B\Sigma_1 C$ is Metzler.
\end{lemma}

The proof of the lemma can be found in \cite{drummond2022aizerman}. Finally we address the positive Aizerman conjectur here. As outlined in the work \cite{drummond2022aizerman}, the Aizerman conjecture is true for positive systems. The authors introduce a theorem that establishes the conditions for the global exponential stability of a positive Lur'e system.
\begin{theorem}[\textbf{Positive Aizerman\cite{drummond2022aizerman}}]\label{the:positiveaizerman}
    Consider a Lur'e-type system \eqref{eq:positiveluresystem} with $B, C \geq 0,$ and $\Sigma_1$ and $\Sigma_2 (\Sigma_1\leq\Sigma_2)$ of appropriate dimension. If $A+B\Sigma_1C$ is Metzler and $A+B\Sigma_2C$ is Hurwitz, then for every nonlinearity $\Phi \in$ Sector $\left[ \Sigma_1,\Sigma_2 \right]$ in the sense of \eqref{eq:mimo sector bound}, the Lur'e system \eqref{eq:positiveluresystem} is globally exponentially stable.
\end{theorem}

This is a significant finding that underscores the viability of Aizerman's conjecture for positive systems. This theorem is instrumental in analyzing the stability of systems controlled by NN. However, a necessary step is to formulate such NN-controlled systems within the Lur'e system framework. The next part aims to precisely define the problem, framing it within the context established by our discussion on positive systems and their stability characteristics.
\subsection{Problem Formulation}
Building upon our discussion and the theoretical framework established, we proceed to define a specific problem involving an LTI system. The system is mathematically characterized as follows:

Consider an LTI system defined by \eqref{eq:generallti}.
For the control policy $u(t)$ consider an output feedback controller $\pi$ in the feedback loop of the system as shown in Fig. \ref{fig:syslook}. The controller $\pi: \mathbb R^p \mapsto \mathbb R^m$,  is realized as a fully connected FFNN with $q$ layers, delineated by:
\begin{subequations}\label{eq:NNcontroller}
\begin{align}
    &\omega^{(0)}(t) = Cx(t),\\
    &\nu^{(i)}(t) = W^{(i)}\omega^{(i-1)}(t)+b^{(i)},i =1,\dots,q\\
    &\omega^{(i)}(t) = \phi^{(i)}(\nu^{(i)}(t)), i =1,\dots,q \label{eq:qthlayer}\\
    &u(t) = W^{(q+1)}\omega^{(q)}(t) + b^{(q+1)},
\end{align}
\end{subequations}
where $\omega^{(i)} \in \mathbb R^{l_i}$ are the output from the $i$th layer with $l_0$ and $l_{q+1}$ being imposed by the input and output size to be $p$ and $m$ respectively.
$\nu^{(i)} \in \mathbb R^{l_i}$ is the vector of the preactivation logits of the $i$th layer.
The operations of each layer are characterized by a weight matrix $W^{(i)} \in \mathbb R^{l_i\times l_{i-1}}$, a bias vector $b^{(i)} \in \mathbb R^{l_i}$, and an activation function $\phi^{(i)}$, applied in an elementwise manner.
\begin{equation}\label{eq:elementwisenonlinearity}\small
\phi^{(i)} (\nu^{(i)}) = \left[\varphi(\nu^{(i)}_1),\varphi(\nu^{(i)}_2),\dots,\varphi(\nu^{(i)}_{l_1})\right]^T,
\end{equation}
where $\varphi: \mathbb R\mapsto \mathbb R$ is the scalar activation function of the NN. 
We assume that the activation function $\phi$ is uniform across all layers of the FFNN. This assumption simplifies the presentation of the behavior of the NN; however,
this assumption can be relaxed with minor modifications to the notation.
We denote the equilibrium state of the system by the set of $(x_*,y_*,u_*)$ where{\small
$$\mathbf{0}_n = Ax_*(t) + B u_*(y_*), y_*(t) = C x_*(t),\text{and\,}u_*(t) = \pi(Cx_*(t)). $$}

Our objective is to study global stability
of the equilibrium state. To achieve this, we plan to transform the closed-loop system, described by \eqref{eq:generallti} and \eqref{eq:NNcontroller}, into a Lur'e-type  system. This transformation enables us to use the positive Aizerman theorem as our tool for analyzing the stability of the system. 

\section{Main Results}
In this section, we take several steps to present our findings on stability assurance of NN-controlled system. Initially, we reconfigure the system into a Lur'e-type framework. Consequently, the need for sector bounding the nonlinear part urges us to find a sector bound for system's nonlinear component. We propose a sector bound based on the weights of the NN and the sector bounds of scalar activation function chosen for NN. These crucial steps set the groundwork for applying the positive Lur'e theorem to analyze system stability.
\subsection{Neural Network controlled system as Lur'e system} 
In this study, we assume the LTI system as in \eqref{eq:generallti} as the linear part of the Lur'e system and consequently include the entire $q$-layer NN denoted by \eqref{eq:NNcontroller} as the nonlinear part. Now, to complete the definition of the Lur'e system, it is required to sector-bound its nonlinear section.

In our effort to sector bound the output of NN, we look for $\Sigma_1$ and $\Sigma_2$ that satisfy the inequality \eqref{eq:mimo sector bound}.
For a positive system ($z\in\mathbb R^n_+$) we should have the following:
\begin{equation*}
     [\Sigma_1]_{m \times n} z_{n \times 1} \leq \pi_{m\times 1}(z) \leq [\Sigma_2]_{m \times n} z_{n \times 1}.
\end{equation*}
Note that the nonlinear activation function, $\phi$, as defined in \eqref{eq:elementwisenonlinearity}, acts elementwise on its argument.
\begin{theorem}\label{the:theorem2}
    Suppose a fully connected FFNN controller $\pi(.)$ with $q$ layers, as defined in \eqref{eq:NNcontroller}, with weights of each layer shown by $W^{(i)}$ and the biases $b^{(i)}$ set to zero. The NN takes $z\in\mathbb R^p_+$ as input and returns $u(z) \in \mathbb R^m.$ Assume identical activation functions for all neurons and assume that the chosen scalar activation function lies in sector $[a_1,a_2],(a_1<a_2)$ and let $c = \max(|a_1|,|a_2|).$
    Define
    \begin{equation}\label{eq:gammas}
    \small
    \Gamma_1 = -c^q\left(\prod_{i=1}^{q+1}\absElem{W^i}\right) \quad \Gamma_2 = c^q\left(\prod_{i=1}^{q+1}\absElem{W^i}\right).
    \end{equation}The defined NN is sector-bounded in the following interval:    \begin{equation}\label{eq:NNsectorbound}
    \small
    \Gamma_1 z \leq \pi(z) \leq \Gamma_2 z.
\end{equation}
\end{theorem}
\textit{Proof:}
The proof has two parts. First, it uses a mathematical induction to establish that the output of the $q$th layer \eqref{eq:qthlayer} is sector-bounded in $-c^q\left(\prod_{i=1}^{q}\absElem{W^i}\right) z \leq \pi(z) \leq c^q\left(\prod_{i=1}^{q}\absElem{W^i}\right)z.$ Then it adds the weights of the output layer $W^{(q+1)}$ to the inequality to establish the theorem.

\textbf{Base Case:}
The sector bounds for the output of the first layer of the NN, defined in \eqref{eq:NNcontroller} with $\omega^{(0)} = z$, are given by:
\begin{equation}\label{eq:ithlayersecor}\small
    -c \absElem{W^{(1)}}z \leq \phi(W^{(1)}z) \leq c \absElem{W^{(1)}}z,
\end{equation}
assuming the scalar activation function $\varphi$ lies within the sector $[a, b]$ with $c = \max(|a|,|b|)$ and $y\in\mathbb R^p_+$. Note that both sides of the inequalities are vectors of size $\mathbb R^{l_1}$ and the inequality is elementwise.
Here a sector bound for the output of first layer is calculated.\\
\textbf{Inductive Step:} 
Assume the sector bounds hold for the output of $i$th layer. Specifically, we have:
\begin{equation}\label{eq:midcalc}
\small
    -c^{i}\left(\prod_{j=1}^{i}\absElem{W^j}\right)z \leq \phi^{(i)}(\dots) \leq c^{i}\left(\prod_{j=1}^{i}\absElem{W^j}\right)z.
\end{equation}
We seek to prove these bounds extend to output of the $i+1$th layer. Through multiplying the output of $i$-th layer by $W^{(i+1)} \in \mathbb R^{l_{i+1} \times l_i}$ we get:
{\small
\begin{align}\label{eq:hava}
  -c^{i}\left(\prod_{j=1}^{i+1}\absElem{W^j}\right)z \leq W^{(i+1)}\phi^{(i)}(\dots) \leq
    c^{i}\left(\prod_{j=1}^{i+1}\absElem{W^j}\right)z,
\end{align}}
with the inequalities being elementwise acting on vectors of size $\mathbb{R}^{l_{i+1}}$.\\
Feeding this preactivation logits to the activation function $\phi^{i+1}$, and given $\phi^{i+1}(.)$ is in sector $[a_1,a_2]$, we obtain:
\begin{align}\label{eq:midcalc1}
    -c\absElem{W^{(i+1)}\phi^{(i)}(\dots)}&\leq \phi^{(i+1)}(W^{(i+1)}\phi^{(i)}\dots) \leq\nonumber \\&
    c\absElem{W^{(i+1)}\phi^{(i)}(\dots)}.
\end{align}
As explained in the notation section, we can use the property $\absElem{W^{(i+1)}\phi^{(i)}} \leq \absElem{W^{(i+1)}}\absElem{\phi^{(i)}}$, and rewrite \eqref{eq:midcalc1} as:
\begin{align}\label{eq:midcalc2}
    -c\absElem{W^{(i+1)}}\absElem{\phi^{(i)}(\dots)}&\leq \phi^{(i+1)}(W^{(i+1)}\phi^{(i)}\dots) \leq\nonumber \\&
    c\absElem{W^{(i+1)}}\absElem{\phi^{(i)}(\dots)}.
\end{align}

From \eqref{eq:midcalc} we deduce:
{\small$
\absElem{\phi^{(i)}(\dots)} \leq \absElem{c^{i}\left(\prod_{j=1}^{i}\absElem{W^j}\right)z}$.}
Note that both $c^i,z \geq 0$ can be moved out of $\absElem{.}$. We use the above inequality in \eqref{eq:midcalc2} and write:
{\small\begin{align*}
    -c^{i+1}\left(\prod_{j=1}^{i+1}\absElem{W^j}\right)z &\leq \phi^{(i+1)}(W^{(i)}\phi^{(i)}(\dots)) \leq \\
    &c^{i+1}\left(\prod_{j=1}^{i+1}\absElem{W^j}\right)z.
\end{align*}}
With the result of induction, the $q$th layer is sector-bounded in $[-c^q\left(\prod_{i=1}^{q}\absElem{W^i}\right) z , c^q\left(\prod_{i=1}^{q}\absElem{W^i}\right)z].$\\
In the second step of the proof, we simply multiply the weights of the output layer $W^{(q+1)}$ to the nonlinearity. As a result of the same operation in \eqref{eq:hava} we obtain:
\begin{equation*}
      \small
   -c^q\left(\prod_{i=1}^{q+1}\absElem{W^i}\right) z \leq \pi(z) \leq c^q\left(\prod_{i=1}^{q+1}\absElem{W^i}\right)z. ~~\blacksquare
\end{equation*}
Given the transformation of the NN-controlled system into a Lur'e configuration \eqref{eq:positiveluresystem}, with sector-bounded nonlinearity, we employ positive Aizerman's theorem (Theorem~\ref{the:positiveaizerman}) for stability analysis. The conditions for stability assurance are succinctly outlined in the following theorem.
\begin{theorem}\label{the:ourtheorem}
Consider a closed-loop system governed by a linear dynamic model described in \eqref{eq:generallti} and controlled by a NN-based policy $u(\cdot) = \pi(Cx)$ as defined in \eqref{eq:NNcontroller}. The NN $\pi(\cdot)$ comprises $q$ layers, with weights of the $i$th layer denoted by $W^{(i)}$, and biases $b^{(i)}$ set to zero. All neurons employ an identical scalar activation function sector bounded within $[a_1, a_2]$, where $c = \max(|a_1|, |a_2|)$. Define the sector bounds $\Gamma_1$ and $\Gamma_2$ as \eqref{eq:gammas}. The combined system dynamics are given by:
\begin{equation}\label{eq:thorem}
\small
\dot x(t) = Ax(t) + B \pi(Cx(t)).
\end{equation}

Assume that $B, C \geq 0$. Additionally, let the system reach equilibrium at $(x_*,y_*, u_*) = \mathbf{0}$. If $A + B \Gamma_1 C$ is Metzler and $A + B \Gamma_2 C$ is Hurwitz, then the closed-loop system exhibits global exponential stability.
\end{theorem}
For the proof of Theorem \ref{the:ourtheorem}, we follow the steps of the proof of Theorem \ref{the:positiveaizerman} as in \cite{drummond2022aizerman}. 

\textit{Proof:}
Given $x(0) \in \mathbb{R}^n_+$ and the NN controller $\pi \in$ Sector $[\Gamma_1, \Gamma_2]$, Lemma \ref{lem:lemma1} guarantees that $x(t) \geq 0$ for all $t \geq 0$. This permits us to reformulate and approximate equation \eqref{eq:thorem} as:
\begin{equation}
\dot{x} = (A + B\Gamma_2C)x + B(\Phi(Cx,t) - \Gamma_2Cx) \leq Mx,
\label{eq:midcalcul1}
\end{equation}
where $M = A + B\Gamma_2C$ is Metzler and Hurwitz. From Fact \ref{fact1}, it follows that there exists a vector $v \in \mathbb{R}_+^n$ with $v > 0$, and a scalar $\epsilon > 0$ such that:
\begin{equation}\label{eq:midcalcul2}\small
v^TM \leq -\epsilon v^T.
\end{equation}
Given that $v$ is strictly positive, we get:
\begin{equation}\label{eq:midcalcul3}
v_m\|z\| \leq v^Tz \leq v_M\|z\|, \quad \forall z \in \mathbb{R}_+^n
\end{equation}
where $v_m$ and $v_M$ are the minimum and maximum elements of $v$, respectively. Simple calculations using \eqref{eq:midcalcul2} and \eqref{eq:midcalcul1} yield the result that for almost all $t \geq 0$:
{\small
\begin{equation*}
    \frac{d}{dt}e^{\epsilon t}v^Tx(t) = \epsilon e^{\epsilon t}v^Tx(t) + e^{\epsilon t}v^T\dot{x}(t) \leq e^{\epsilon t}(\epsilon v^T + v^TM)x(t) \leq 0.
\end{equation*}}
Since $t \mapsto e^{\epsilon t}v^Tx(t)$ is nonnegative, \eqref{eq:midcalcul3} yields:
\begin{equation*}
e^{\epsilon t}v_m\|x(t)\| \leq e^{\epsilon t}v^Tx(t) \leq v^Tx(0) \leq v_M\|x(0)\|, \forall t \geq 0
\end{equation*}
from which the global exponential stability is deduced. \hfill $\blacksquare$

\section{Examples}
In this section, we provide an example to demonstrate the applicability of our theorem. Consider the following system:
{\small
\begin{align}\nonumber
\dot x(t) = \begin{bmatrix}
    -5 & 1 \\
    3 & -5
\end{bmatrix}x(t) + \begin{bmatrix}
    0.5 \\
    1
\end{bmatrix}u(t), \quad y(t) = I_{2\times2}x(t). \nonumber
\end{align}}

\begin{figure}[t]
    \centering
            \begin{subfigure}{.25\textwidth}
                \centering
                \includegraphics[width=1\linewidth]{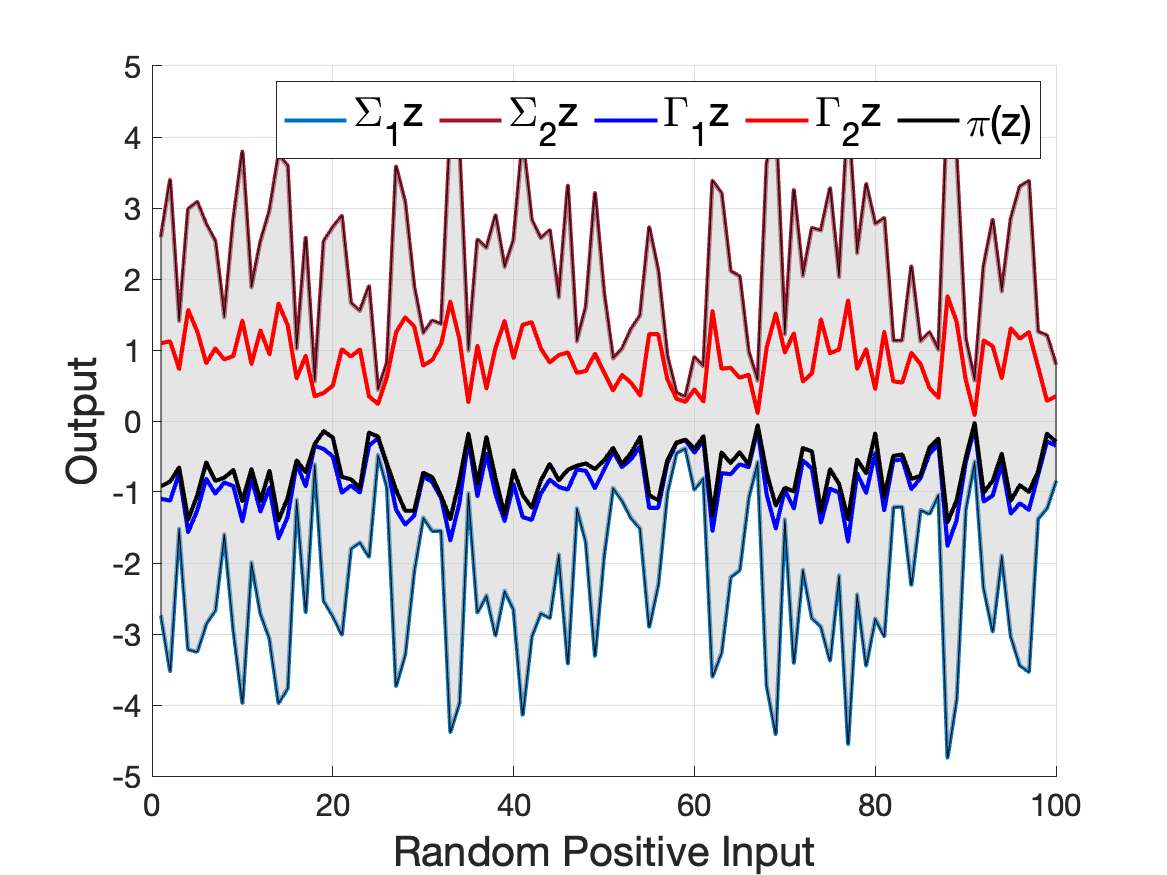}
                \caption{\small \centering 3-layer NN trained on LQR with $Q =$ diag$(0.3,3)$, $R = 0.3$}
                \label{fig:tightbound}
            \end{subfigure}%
            \begin{subfigure}{.25\textwidth} %
                \centering
                \includegraphics[width=1\linewidth]{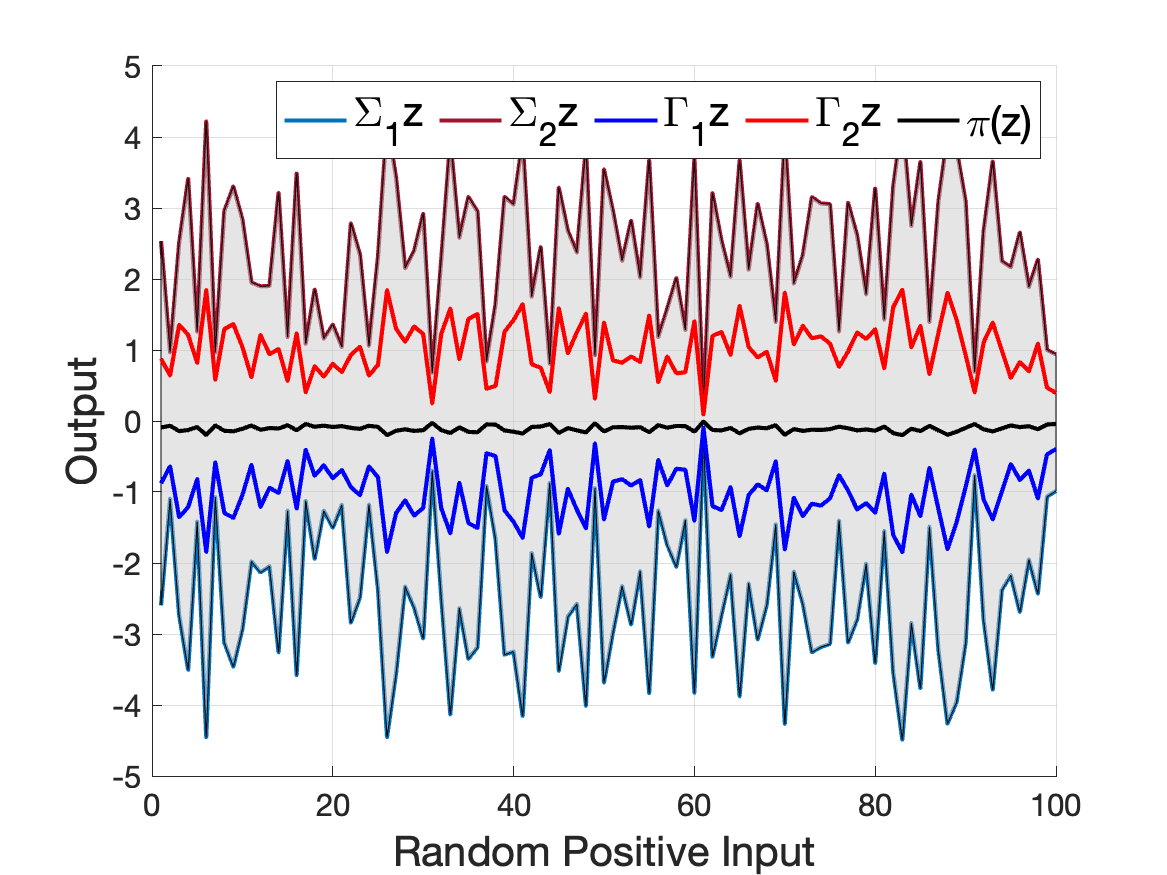}
                \caption{\small\centering 4-layer NN trained on LQR with $Q = \mathbf I_2$ and $R = 1$.}
                \label{fig:widebound}
            \end{subfigure}
    \caption{\small The sector bounds vs NN output for two different NNs.}
        \vspace{-.3cm}
    \label{fig:Sector}
\end{figure}

The controller is parameterized by an fully connected FFNN with $4$ layers, containing $10,15,15,1$ neurons in each hidden layer. The activation function $\tanh$, sector-bounded in $[0,1]$, is selected for all neurons. Biases are set to zero, therefore, $(x_*,y_*,u_*) = \mathbf{0}$ is an equilibrium point. NN controller is trained to mimic an LQR controller with $Q=\mathbf{I}_2$ and $R=1$. Upon completion of training, the NN $\pi$ serves as a deterministic controller $u(t) = \pi(Cx)$. As the result of training, the weights of the NN yield $-\Gamma_1 = \Gamma_2 = [2.75,1.47]$. In Fig. \ref{fig:Sector}, the validity of the sector bounds defined by \eqref{eq:NNsectorbound} is demonstrated. Fig. \ref{fig:Sector} shows the NN output for a set of $100$ random inputs to lie between the functions $\Gamma_1z$ and $\Gamma_2z$. As demonstrated in Fig. \ref{fig:Sector}, the bounds obtained by the widest possible \(\Sigma_1\) and \(\Sigma_2\) in Theorem \ref{the:positiveaizerman} encompass those obtained by \(\Gamma_1\) and \(\Gamma_2\) for two different NNs.
As seen in Fig. \ref{fig:tightbound} our obtained bound can be tight, depending on the neural network's parameters.

\begin{figure}[ht] 
    \centering
    \resizebox{\columnwidth}{!}{
            \begin{subfigure}{.35\textwidth} 
                \centering
                \includegraphics[width=\linewidth]{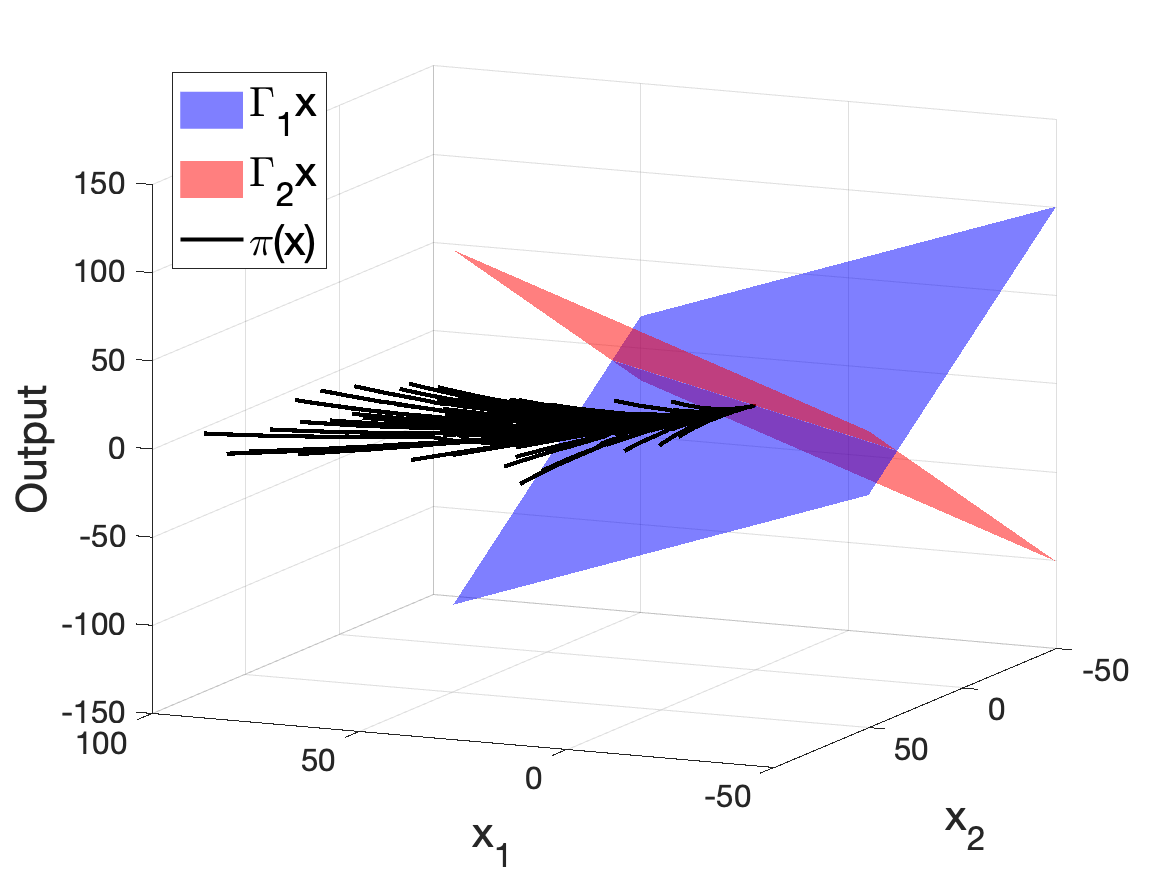} 
                \caption{\small Trajectories of input and states.}
                \label{fig:sectortrajectory}
            \end{subfigure}%
            \begin{subfigure}{.35\textwidth} 
                \centering
                \includegraphics[width=\linewidth]{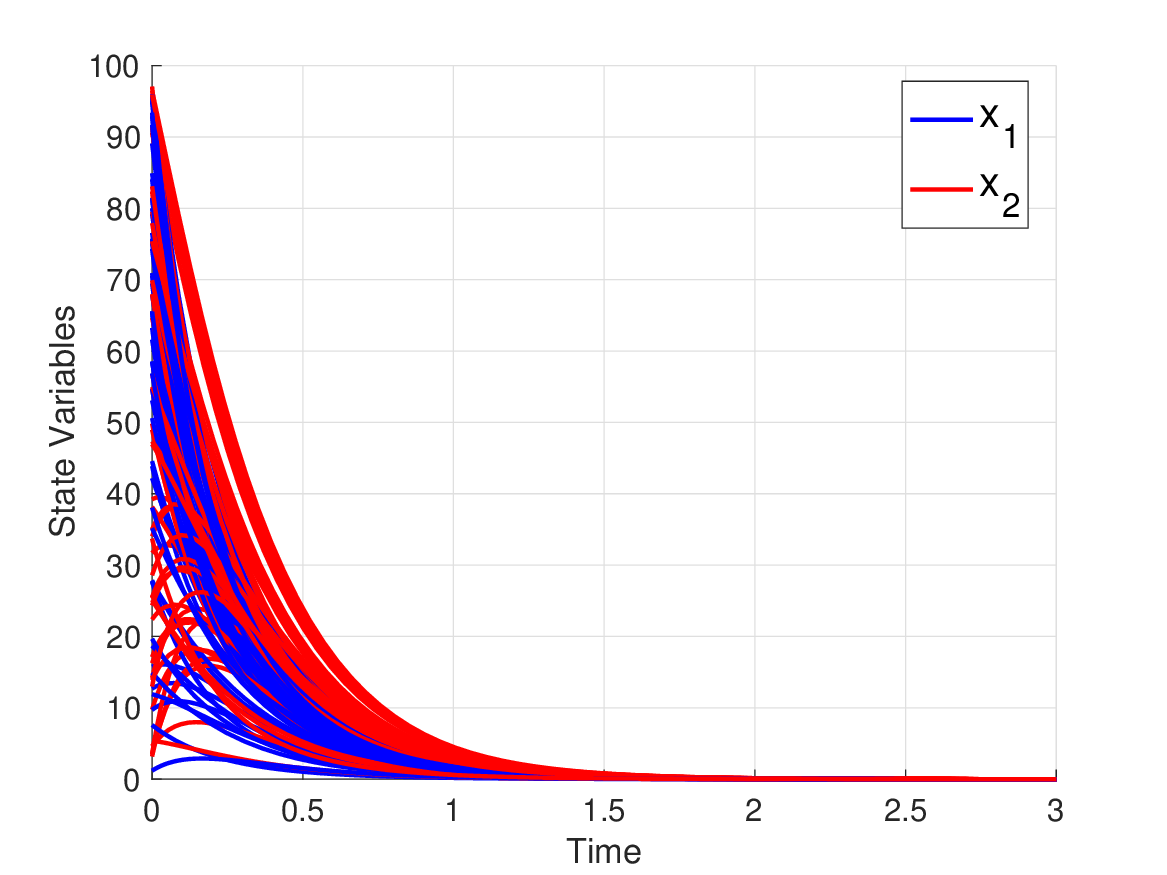}
                \caption{\small Trajectories of states.}
                \label{fig:states}
            \end{subfigure}}
    \caption{\small Trajectory of system for $50$ random initial conditions.}
        \vspace{-.75cm}
    \label{fig:trajectoriess}
\end{figure}

Now, given that {\small$A+B\Gamma_1C = \begin{bmatrix}
-5.41 & 0.40\\2.17& -6.18\end{bmatrix}$} is Metzler and $A+B\Gamma_2C$ is Hurwitz with eigenvalues $(-6.69,-1.70)$, we anticipate the global exponential stability of the closed-loop system. Figs. \ref{fig:sectortrajectory} and \ref{fig:states} demonstrate stability. In Fig. \ref{fig:sectortrajectory}, the trajectories of the input of the system are shown for $50$ random initial conditions. All of the trajectories remain inside the sector and finally converge to equilibrium. Furthermore, Fig. \ref{fig:states} shows the trajectories of the states of the system, for the same random initial conditions. The exponential stability is foreseeable from Fig. \ref{fig:states}.
In order to demonstrate the scalability of the approach, we compared the average run-time of our method against the well-known IQC method presented in \cite{yin2021stability}. Both codes were run on a same benchmark with Matlab. The result is shown in the first row of Table \ref{tab:benchmark}. In the second and third columns, we compared our sector bounds with one of the conservative bounds from the literature \cite{szegedy2013intriguing} to assess the level of conservatism in our sector bounds. As observed, our sector bounds exhibit better performance. Furthermore, since our sector bound highly depends on each entry of weight matrices, we can leverage this characteristic to train a NN with an additional layer and almost the same sector bounds. An example of such capability is shown in Table \ref{tab:benchmark}. The addition of a layer had minimal impact on our bound, while increased the other conservative bound.

\begin{table}[h]
\centering
\caption{\footnotesize Comparison of computation time and bounds. $10/10/1$ denotes a 3-layer NN with $10, 10, 1$ neurons, respectively.}
\footnotesize
\resizebox{\columnwidth}{!}{
\begin{tabular}{|c|c|c|c|}
\hline
\textbf{Method} & \textbf{Network Architecture} & \textbf{Computation Time (s)} & \textbf{Bounds} \\
\hline
\textbf{Our method} & $10/10/1$ & $2.5 \times 10^{-5}$ & $\pm[2.65,1.61]$ \\
\hline
\textbf{Our method} & $10/15/15/1$ & $2.6 \times 10^{-5}$ & $\pm[2.75,1.47]$ \\
\hline
\textbf{IQC method$^\star$\cite{yin2021stability}} & $10/10/1$ & $0.68$ & ------ \\
\hline
\textbf{Product of Norms $^{\star \star}$\cite{szegedy2013intriguing}} & $10/10/1$ & ------ & $5.83$ \\
\hline
\textbf{Product of Norms $^{\star \star}$ \cite{szegedy2013intriguing}} & $10/15/15/1$ & ------ & $6.45$ \\
\hline
\end{tabular}}
\label{tab:benchmark}
\begin{tablenotes} 
\item $^{\star}$~IQC does not explicitly provide bounds for entire NN. Therefore, bounds are not reported. $^{\star \star}$~Not a method to verify stability; only for comparing bounds. Therefore, computation time is not reported.
\end{tablenotes}
\end{table}

\section{Discussion \& Conclusion}
In this section, we discuss our results and conclude the paper. Our analysis uncovers several critical observations. 
We expect that increasing the number of layers in the neural network will naturally expand its sector bounds, potentially making it more challenging to verify the stability of a broader range of systems. However, while this is generally the case, the structure of our sector bounds provides some flexibility to offset this expansion, as illustrated in Table \ref{tab:benchmark}.
Moreover, the addition of more layers does not add complexity or difficulty to our verification method.

To contextualize our findings, we compare them with previous studies in the field. We note that most studies focus on using Lyapunov functions and solving Linear Matrix Inequalities (LMIs) for stability. These methods are less practical for large systems due to their complexity. Our approach, based on the Aizerman conjecture, offers a simpler alternative that results in a huge improvement of run-time and scalability to larger systems.

While interpreting the results, it is essential to acknowledge certain limitations. A notable instance is that our bounds can become conservative in some cases. Two examples are shown in Fig. \ref{fig:Sector} for both conservative and quite precise bounds. One way to overcome this limitation is to adopt local sector bounds, as explained in \cite{yin2021stability}. This essentially means restricting the inputs to the activation functions, or in another word, reducing the level of stability to local stability of the closed-loop system.

Another limitation is the restriction of our sector bound in handling NN with biases. This limitation carries over to our analysis. While our sector bounds can be extended to include biases with some modifications, detailing this would exceed the scope of the current paper. We plan to explore this extension in future work. 
Another less restrictive assumption is that $C,B\geq0$ as the state transition matrix $A$ is not restricted in the positive Aizerman theorem. Matrix $A$ does not need to be Metzler, and therefore the LTI system does not need to be positive. A verification method based on the positive Aizerman theorem only requires the combined LTI system and NN controller to be positive. 

This study encourages the exploration of basic theorems like the Aizerman and Kalman conjectures for verification of NNs. It opens the door for further investigations into refining sector bounds for NNs to enhance stability analysis. The potential for investigating local sector bounds and the local stability of systems, and examining other NN architectures, such as recurrent neural networks, within the framework of the positive Aizerman theorem presents an intriguing avenue for further study.

\begin{spacing}{.81}
\bibliography{References} 
\end{spacing}
\end{document}